\documentstyle[prl,aps,preprint,epsf]{revtex}

\begin{document}
\newcommand{\be}{\begin{equation}}
\newcommand{\ee}{\end{equation}}
\newcommand{\bq}{\begin{eqnarray}}
\newcommand{\eq}{\end{eqnarray}}
\newcommand{\Sc}{Schr\"odinger\,\,}
\newcommand{\Sp}{\,\,\,\,\,\,\,\,\,\,\,\,\,}
\newcommand{\no}{\nonumber\\}
\newcommand{\tr}{\text{tr}}
\newcommand{\p}{\partial}
\newcommand{\la}{\lambda}
\newcommand{\G}{{\cal G}}
\newcommand{\D}{{\cal D}}
\newcommand{\W}{{\bf W}}
\newcommand{\de}{\delta}
\newcommand{\al}{\alpha}
\newcommand{\bi}{\beta}
\newcommand{\ga}{\gamma}
\newcommand{\ep}{\epsilon}
\newcommand{\vep}{\varepsilon}
\newcommand{\th}{\theta}
\newcommand{\om}{\omega}
\newcommand{\J}{{\cal J}}
\newcommand{\pr}{\prime}
\newcommand{\ka}{\kappa}
\newcommand{\TH}{\mbox{\boldmath${\theta}$}}
\newcommand{\DE}{\mbox{\boldmath${\delta}$}}

\setcounter{page}{0}
\def\footnoterule{\kern-3pt \hrule width\hsize \kern3pt}
\tighten
\title{
$(1+1)$-Dimensional $SU(N)$ Static Sources in $E$ and $A$ Representations
}
\author{Jiannis Pachos
\footnote{Email address: {\tt pachos@ctp.mit.edu}}
}
\address{Center for Theoretical Physics \\
Massachusetts Institute of Technology \\
Cambridge, Massachusetts 02139 \\
{~}}

\date{MIT-CTP-2698, January 1998}
\maketitle

\thispagestyle{empty}

\begin{abstract}

Here is presented a detailed work on the $(1+1)$ dimensional $SU(N)$ Yang-Mills theory with static sources. By studying the structure of the $SU(N)$ group and of the Gauss' law we construct in the electric representation the appropriate wave functionals, which are simultaneously eigenstates of the Gauss' operator and of the Hamiltonian. The Fourier transformation between the A- and the E-representations connecting the Wilson line and a superposition of our solutions is given. 

\end{abstract}

\vspace*{\fill}

\newpage

\section{Introduction}

In this work the $(1+1)$-dimensional Yang-Mills theory with static sources in the temporal gauge is studied. As it is well known, in the coordinate A-representation the solutions of the Gauss' law constraint and the Hamiltonian eigenvalue problem are given in terms of the Wilson line of the vector potential $A$. The equivalent problems in the momentum E-representation (\cite{Jack},\cite{Freedman}) are studied here for the $SU(N)$ group as an extension of a previous work for $SU(2)$, \cite{AJ}. In addition, another functional is presented, which apart from an irrelevant constant is a superposition of the solutions in \cite{AJ}. This wave functional is explicitly Fourier transformable, giving the Wilson line. Using these functionals the confining potential between the static sources is derived and the connection of this $(1+1)$ dimensional quantum problem with the equivalent classical one is studied. These constructions are made possible after an appropriate decomposition of the $SU(N)$ group with respect to its Cartan sub-algebra. The Gauss' law undergoes a similar decomposition. 

\section{$SU(N)$ group}

In this section we shall describe some properties of the $SU(N)$ group and its corresponding algebra. In particular we are interested in the Cartan decomposition and the consequences this has. The group used here is $G=SU(N)$ with algebra ${\cal G}= {\it su}(N)$. Let $H$ be the Cartan sub-group generated from the ${\cal H}$ Cartan sub-algebra of $su(N)$. We define $T_a$ to be the $N^2-1$ matrix generators of $G$ belonging to some matrix representation $r$ and satisfying the commutation relation $[T^a,T^b]= if_{abc}T^c$ and the normalization condition $\tr T^a T^b=c\de _{ab}$. An element, $g$, of $G$ in this representation can be written in terms of $N^2-1$ parameters, $\omega _a$, as $g=e^{\omega_aT_a}$, with $\om_a=\om_a(\chi_1,..,\chi_{N^2-1})$ depending on $N^2-1$ angles, $\chi$. We can also define the differential generators ${\cal J}_a(\chi(x))$ in terms of functional differentiation with respect to these angles $\chi(x)$, as ${\cal J}_a(\chi(x))=L_{ab}(\chi(x)) {\de \over \de \chi_a(x)}$, where the $(N^2-1)\times (N^2-1)$ matrices $L_{ab}$ are invertible. 

Without loss of generality we can ask that $g$ be parameterized in terms of the angles $\chi(x)$, so the following relation holds.
\be
{\cal J}_a(x) g =-T_a g \,\, .
\label{left}
\ee
Also, we can demand
\be
{\cal J}^R_a(x)g=gT_a \,\, ,
\label{right}
\ee
where ${\cal J}^R_a(x)$ are the ``right'' differential operators connected with the ``left'' ones by a rotation belonging in the adjoint representation. They can be written as ${\cal J}^R_a(\chi(x))=R_{ab}(\chi(x))$ ${\de \over \de \chi_a(x)}$.

We arrange the $N^2-1$ generators $T_a$ so that the Cartan ones are the first $N-1$, i.e. $T_k$, for $k=1,...,N-1$. As they commute with each other we can take them to be diagonal with elements given by $(T_k)_{i j}=\de_{i j}f^k(i)$, for some functions, $f^k$. From (\ref{left}) we expect that there should be $N-1$ commuting differential operators ${\cal J}_k$, which can be expressed as a differentiation with respect to single angles, $\phi_k$, i.e. ${\cal J}_k=-i{\de \over \de \phi_k}$. The same is true for the right operators, which can be written as ${\cal J}^R_k=-i{\de \over \de \overline{\phi}_k}$ with respect to some { \it different} angles, $\bar \phi _k$. The ``diagonalized'' ${\cal J}^R_k$ are referred in the literature as $P_k$. The complete set of angles, $\chi$, is decomposed as $\chi=(\phi,\bar \phi, \th)$.

In view of these diagonalizations we can express $g$ as
\be
g_{i j}=e^{if^k(i)\phi_k} e^{-if^k(j)\bar \phi ^k} \tilde g _{i j}(\th)=(h(\phi^k)\tilde g(\th) h(\bar \phi^k))_{i j} \,\, ,
\label{diag}
\ee
where $\tilde g(\th)$ does not dependent on $\phi$ or $\bar \phi$, but on the remaining $(N-1)^2$, $\th$ angles. 

With the above decomposition the integration measure of the group, $\D g$, becomes, $[\D\phi]$ $[\D \bar \phi] [\D \th] J(\th)$, with $[\D\phi]=\D\phi_1..\D\phi_{N-1}$, $[\D \bar \phi]=\D\bar \phi_1..\D \bar \phi_{N-1}$, $[\D \th]=\D \th_1..$ $\D\th_{(N-1)^2}$, while the Jacobian, $J$, depends only on the $(N-1)^2$, $\th$, angles. This property allows in the following the straightforward evaluation of the integrals with the construction of delta functions.

For the $SU(2)$ group, the above decomposition is obvious, as its general element, $g$, can be parameterized as $g(\al,\bi,\gamma)=e^{-i\al \sigma_3/2}e^{-i\bi \sigma_2/2}e^{-i\gamma \sigma_3/2}$. The $SU(3)$ case can be retrieved from reference \cite{Byrd} with simple modifications.

\section{Gauss' law}

The constraint of the Gauss' law in the A-representation implies that the constructed wave functional should be invariant under $SU(N)$ gauge transformations. In the E- representation this constraint forces the wave functional to transform in the following way (see \cite{Jackiw}, \cite{Nair})
\be
\Psi[E^U]=e^{-{1 \over c} \tr \int dx E U^{-1}U^{\prime}}\Psi[E] \,\,  .
\label{tr}
\ee
The rotated vector, $E^U$, is given by $UEU^{-1}$. $E=E^aT^a$ depends on $N(N-1)$ angles, which parameterize the coset $G/H$, and on $N-1$ amplitudes. Together these variables are $N^2-1$, equal to the independent variables of the $SU(N)$ group. As the Gauss' law in E-space is a condition imposed on the wave functional, $\Psi[E]$, when rotations of $E$ are considered, it needs only $N(N-1)$ independent equations to define its transformation properties in terms of the corresponding angles.

For the electric Schr\"odinger representation the Gauss' operator is given by
\be
G_a(E(x))=\p _xE_a(x)-if_{abc}E_b(x) {\de  \over \de E_c(x)}=\p _xE_a(x) +J_a(x) \,\,  .
\ee
$a$ runs from 1 to $N^2-1$. If we derive the Gauss' law from the infinitesimal form of the gauge transformation (\ref{tr}), with respect to the $N(N-1)$ angles, we shall obtain $N(N-1)$ equations. To consider the remaining $N-1$ equations let us observe the commutation relation of the operators, $G_a$,
\be
[G_i(x),G_k(y)]\Psi[E]=if_{ikj} G_j(x)\Psi[E] \de(x-y) \,\,  .
\ee
If we restrict $i$ to run over the $N(N-1)$ parameters of $G/H$ and also $k$ to run over the $N-1$ parameters of $H$, then $j$ will run over the $G/H$ indices, in parallel with the corresponding generators $T^a$, when satisfying the Cartan-Weyl set of commutation relations. This results that $G_i(x)G_k(y)\Psi[E]=0$. A stronger requirement is imposed from the Gauss' law asking that $G_k(y) \Psi[E]=(E_k^{\pr}+J_k)\Psi[E]=0$. It is always possible to find a coordinate frame where $J_k \Psi[E]=0$. Consequently, these $N-1$ equations will constraint the gauge amplitudes of the electric field via delta functions in $\Psi[E]$, as we shall see in the following example.

For the $(1+1)$-dimensional $SU(2)$ case we can write the Gauss' law in spherical coordinates $(|E|,E_{\th},E_{\phi})$, rather than Cartesian $(E_1,E_2, E_3)$, as
\bq
&|E|^{\prime}\Psi[E]=0&
\\ \no
&(|E|E_{\phi}^{\prime} +i {1 \over \sin E_{\th} } { \de \over \de E_{\th}}) \Psi[E]=0&
\\ \no 
&(|E| E_{\th}^{\prime} \sin E_{\th} -i{\de \over \de E_{\phi}} ) \Psi[E]=0&
\eq
where the parameters $E_{\th}$ and $E_{\phi}$ are given from $E_1\!=\!|E|\cos E_{\phi} \sin E_{\th}$, $E_2\!=\! |E|\sin E_{\phi} \sin E_{\th}$ and $E_3\!=\!|E|\cos E_{\th}$. While the first equation demands that the wave functional contains a delta function which constrains the electric field to have a constant amplitude, $|E|$, with respect to $x$, the remaining equations determine how the wave functional transforms when the vector $E$ is rotated. The delta functions will be derived in a following section for the $SU(N)$ case as well as the part of $\Psi[E]$ that provides the proper transformation properties. 

\section{Solving the Gauss' law with sources}

We want to solve the modified Gauss' law, $G_a(x)\Psi[E]=\Psi[E]T_a \de(x-x_0)-T_a\Psi[E] \de(x-x_1)$, for the general $SU(N)$ group. It represents the vacuum with two static conjugate sources embedded at the points $x_0$ and $x_1$. The vacuum case is retrieved by taking $x_0$ to minus infinity and $x_1$ to plus infinity. A functional, $\Psi[E]$, which satisfies this relation should transform like
\be
\Psi[E^U]=e^{-{1 \over c} \tr \int dx E U^{-1}(x) U^{\prime}(x)}U(x_1)\Psi[E]U^{-1}(x_0) \,\,  .
\label{ggg}
\ee
Clearly the functional 
\be
\Psi[E]=\int \D u e^{-{1 \over c} \tr \int E u^{\prime} u ^{-1}}u(x_1)u^{-1}(x_0)
\label{nair}
\ee 
transformations in this way, so it is a solution of the modified Gauss' law. There could be a constant matrix between $u(x_1)$ and $u^{-1}(x_0)$ without spoiling the transformation property (\ref{ggg}), as $u$ transforms like $u\rightarrow Uu$ under a $U \epsilon \, SU(N)$ transformation. This result specialized in the vacuum case is studied in \cite{Nair}. 

\section{Fourier Transformation}

The Fourier transformation is widely used in the case of functions. For functionals we have the obstacle that we do not know how to performs functional integrations apart from a handful of cases. Even if the functional Fourier transform is formally well defined, most of the times it cannot be applied simply because we do not know the answer of the functional integration it includes. However, the special case of the integrations of delta functions can be easily performed. Here we shall use a proper generalization of the delta functions for the functional space to be the product of simple ones at each point, $x$, i.e. $\prod_x \de (f(x))$. It can also be defined as
\be
\prod_x \de (f(x)) \equiv \int \D a e^{i \int a(x) f(x)}
\ee
where overall infinite constants have been neglected. 

We can define a Fourier transformation for the functional space as
\be
\widetilde \Psi[A] = \int \D E e^{i \int E_aA_a}\Psi[E]
\ee
where $A_a=-i(\bar g^{\pr} \bar g^{-1})_a$.
Now, we shall show that the proposed functional (\ref{nair}) has the Wilson line as its Fourier transform and via versa. Let us perform the following steps
\bq
&&
\int \D E e^{i \int E_a A_a} \int \D u e^{ -\int E_a (u^{\pr}u^{-1})_a} u(x_1)u^{-1}(x_0)=
\no \no
&&
\int \D u \prod_x \de (A_a +i(u^{\pr}u^{-1})_a) u(x_1)u^{-1}(x_0)\equiv \widetilde \Psi[A] \,\,  .
\label{arep}
\eq
With a change of variables $u \rightarrow \bar g u$ and by bearing in mind the rotational invariance of the integration measure we obtain
\be
\widetilde \Psi[A]=\int \D u \prod_x \de(i(u^{\pr} u^{-1})_a)\bar g(x_1)u(x_1)u^{-1}(x_0) \bar g^{-1}(x_0)
\ee
The delta function enforces $u$ to be a constant matrix for any $x$, so eventually
\be
\widetilde \Psi[A]=\bar g (x_1) \bar g^{-1}(x_0)=Pe^{i \int^{x_1}_{x_0}A(x)dx}\,\, .
\label{line}
\ee
This is the Wilson line of the vector potential along a line connecting the points $x_0$ and $x_1$. We have defined $\widetilde \Psi[A]$ from relation (\ref{arep}) because this is the form which is obvious how to Fourier transform back to the E-representation, in contrast to the Wilson line given in (\ref{line}).

\section{More - Special - solutions of the gauss' law}

Here we shall solve the Gauss' law with sources, by making an ansatz for the wave functional rather than requiring the proper transformation properties. We suggest as solutions of the $SU(N)$ Gauss' law a generalization of the ones given in \cite{AJ} for the simple case of $SU(2)$. Hence, the functional is taken to have the form
\bq
\Psi[E]&=& e^{-{1 \over c} \tr \int E g^{\pr}g^{-1}}\W(x_1)\W^{+}(x_0)
\no \no
&&
\times
\prod_x \de(k_1(x)+\tilde k_1 \TH(x_0,x_1)-\rho_1)... \prod_x \de(k_{N-1}(x)+\tilde k_{N-1} \TH(x_0,x_1)-\rho_{N-1})
\label{ours}
\eq
where $E=gKg^{-1}$, \TH$(x_0,x_1)\equiv \theta(x-x_0)-\theta(x-x_1)$, $\W$ is a column matrix and $\tilde k$ and $\rho$ are constants. The phase, $e^{i\Omega}$, is borrowed from the vacuum functional, so that the Gauss' law is satisfied for every $x \neq x_0$, $x_1$. The column, $\W$, has entries functions of the field and its purpose is to combine the matrix representation of the sources with the differential representation of the angular momentum operators in the Gauss' operator. The delta functions, $\prod_x \de(k_i(x)-\tilde k_i \TH(x_0,x_1)-\rho_i)$, provide the necessary spatial delta functions on the left hand side, to compensate the ones of the sources on the right hand side. 

Substituting it in the Gauss' law we find after some algebra that the crucial equation which has to be satisfied at $x_1$ is (see \cite{AJ} for more details)
\be
{1 \over c} \W \tr \Big(\tilde K g^{-1} (T_a g + J_a g)\Big)+T_a \W +J_a \W =0 \,\,  .
\ee
for $K(x) \equiv \tilde K \TH(x_0,x_1) + P$, while a conjugate one has to be satisfied at $x_0$. We can isolate the derivatives referring to the Cartan angles, $\phi$, in ${\cal J}_a$ as
\be
\J _a = J_a +\sum_{k=1}^{N-1}\Gamma^k_a\left( -i {\p \over \p \bar \phi^k} \right)=J_a +\sum_{k=1}^{N-1}\Gamma^k_aP_k
\ee
for $\Gamma^k_a$ some functions of the angles. As $\tilde K$ is a Hermitian matrix it can be diagonalized and written as a linear combination of the Cartan generators, $\tilde K=\sum_{k=1}^{n-1} \tilde k_kT_k$. In addition we can choose $g$ to satisfy the diagonalization given in (\ref{diag}), so that finally we obtain
\be
-\W\sum_{k=1}^{N-1} \tilde k_k \Gamma_a^k +(T_a \W +J_a \W) =0
\ee
A solution of this equation for $\W$ would be an $s$-column of an $SU(N)$ element, $g$, satisfying also the diagonalization condition (\ref{diag}), so that the above equation for independent functions, $\Gamma_a^k$, finally results
\be
\tilde k_k=-f^k(s)\,\,\,\,\text{for any $k=1,..,N-1$ and for $s$ constant}. 
\label{meas}
\ee

In conclusion, $\W$ should be a column of a diagonalized element, $g$, of $SU(N)$ and the amplitude of $E$ should be restricted by the above delta functions, with $\tilde k_k$ completely defined from (\ref{meas}). In order to obtain finite energy eigenvalues from these solutions we need to choose for $\W(x_0)$ and $\W(x_1)$ the same column, $s$, of $g$, defined at the appropriate points. 

\section{Relationship Between the E-Representation Solutions}

In the following we shall see how the solutions we derived in the E-representation with the two different methods are connected. To demonstrate it we start from (\ref{nair}) and we aim to extract the solutions given in the previous section. As has been shown in \cite{AJ} the superposition of the functionals with different $s$ has the same energy eigenstate as the Wilson line. So we expect that the functional in (\ref{nair}), which has been shown to be the Fourier transform of the Wilson line, will reduce to this superposition. In particular if $E=gKg^{-1}$ we can write (\ref{nair}) as 
\be
\Psi[E]=e^{- {1 \over c} \tr \int E g^{\pr} g^{-1}} g(x_1) \left[ \int \D u e^{- {1 \over c} \tr \int K u^{\pr} u^{-1}} u(x_1) u^{-1} (x_0) \right] g^{-1} (x_0)\,\, .
\ee
Let us introduce the following decomposition of the identity into projection operators, $P^i$, as
\be
{\bf 1} =\sum_{i=1}^r P^i\,\, , \,\,\,\,\text{with}\,\,\,\,P^iP^i=P^i\,\,\,\, \text{(no sum in $i$)}, \,\,\,\,\text{and}\,\,\,\, P^iP^j=0 \,\,\,\,\text{for} \,\,\,\, i \neq j\,\, ,
\ee
with $P^i$ being a diagonal matrix with zeros on the diagonal except on the $i$-th place. We insert two decomposition operators, one on the left of $g(x_1)$ and the other on the right of $g^{-1}(x_0)$ as in the following
\be
\Psi[E]=e^{- {1 \over c} \tr \int E g^{\pr} g^{-1}} \sum_{i,j=1}^{r} g(x_1) \left[ \int \D u e^{- {1 \over c} \tr \int K u^{\pr} u^{-1}} P^i u(x_1) u^{-1} (x_0) P^j \right] g^{-1} (x_0) 
\label{into}
\ee
As a result each term in the double sum depends on the $i$-th row of $u(x_1)$ and on the $i$-th column of $g(x_1)$ as well as on the $j$-th column of $u^{-1}(x_0)$ and on the $j$-th row of $g^{-1}(x_0)$. This property is necessary for having the $g$'s combining properly with the first exponential to satisfy the Gauss' law, as in the previous section, while the outcome with the $u$'s is used to extract the delta functions for the amplitude of the electric field, as we shall see in the following. 

We can perform the integrations with respect to the $ \phi$ angles of $u$, so that the relevant factor becomes
\bq
&&
\int \D  \phi_1 ... \D \phi_{N-1} e^{i\int k^{\pr}_k(x) \phi_k} \sum_{i,j} u_{i\ga}(x_1) u^{-1} _{\ga j}(x_0)=
\no \no
&&
\sum_{i,j} \prod_x \de \Big(k^{\pr}_1(x)+f^1(i)\de(x-x_1) - f^1(j) \de (x-x_0) \Big) ... 
\no \no 
&&
\,\,\,\,\,\,\,\,\prod_x \de \Big( k^{\pr}_{N-1}(x)+f^{N-1}(i)\de(x-x_1) - f^{N-1}(j) \de (x-x_0) \Big) \,\, \bar u _{i \ga}(x_1) \bar u ^{-1} _{\ga j }(x_0) \simeq
\no \no 
&&
\sum_{i,j} \prod_x \de \Big( k_1(x)+f^1(i)\th(x-x_1) - f^1(j) \th (x-x_0)-\rho_1 \Big) ...
\no \no 
&&
\,\,\,\,\,\,\,\,\prod_x \de \Big( k_{N-1}(x)+f^{N-1}(i)\th(x-x_1) - f^{N-1}(j) \th (x-x_0)-\rho_{N-1}\Big)\,\,  \bar u _{i \ga}(x_1) \bar u ^{-1} _{\ga j }(x_0)
\eq
For the second equality we have substituted the delta function condition that the derivative of a function being zero with the condition that the function being a constant, by changing the appropriate arguments. Now we perform the integration with respect the $\bar \phi^k$ variables from $\bar u =\tilde u(\th) h(\bar \phi)$, where we shall extract similar delta functions, which include the $\theta$ variables. They will enforce $\tilde u$ to be equal to the identity, giving eventually
\bq
\sum_{i} \prod_x \de \Big( k_1(x)-f^1(i)\TH(x_0,x_1) -&\rho_1& \Big) ...
\no\no
\prod_x \de \Big( k_{N-1}(x)&-&f^{N-1}(i)\TH(x_0,x_1) -\rho_{N-1}\Big) 
\eq
The above expression, substituted back in (\ref{into}), will give a superposition of the solutions found in the previous section. This conclusion proves the conjecture made in \cite{AJ} for the connection of the Wilson line with this superposition.

\section{Potential of the two static sources}

Now we shall calculate the energy eigenvalues the proposed functionals give. The potential of two static sources (source - anti-source) is
\be
V={1\over 2} \int \! dx \,\,\, \tr \! \int \! \D E \, \Psi ^+ [E] E^a(x)E^a(x) \Psi[E]
\ee 
for $\Psi[E]$ given from (\ref{ours}), where the constant numbers $\tilde k_1, ..., \tilde k_{N-1}$ of $\Psi[E]$ are given by $-f^k(s)$ for $k=1,...,N-1$ and for fixed $s$. To minimize the energy we should take $\rho_1=...=\rho_{N-1}=0$. Then 
\be
V={N+1 \over 2} \sum_{i=1}^{N-1} f^i(s)^2 (x_1-x_0)
\ee
where the factor $N+1$ has been included for the $N+1$ different orientations the Cartan sub-algebra could have. We can calculate the potential for the average in $s$ of these solutions. Easily, we deduce that the potential is proportional to the quadratic Casimir operator, $C_2=c(N^2-1)/r$, that is
\be
V={1 \over 2} C_2 (x_1-x_0) \,\, ,
\ee
where $C_2$ is given from $T^aT^a=C_2 {\bf 1}$. The same result can be derived in the A-representation using the Wilson line solution.

\section{Classical Reduction}

As we know the $(1+1)$-dimensional Yang-Mills problem with sources is not dynamical. The Hamiltonian operator in the electric representation is multiplicative and does not introduce any dynamics as there are no transverse degrees of freedom. Moreover, the Gauss' law defines completely the wave functional. It chooses a special configuration for the electric field which is imposed with the presence of the functional delta functions in the wave functional. This makes the quantum problem similar to a classical one. 
The classical Gauss' law with conjugate static sources is 
\be
E^{\prime}_a-f_{abc}E_bA_c=Q_a \de(x-x_0)-Q_a^{+} \de(x-x_1) \,\, ,
\ee
where $\rho_a(x)=Q_a\de(x-x_0)-Q_a^{+} \de(x-x_1)$ is the charge density of the static sources at the points $x_0$ and $x_1$. From the above equation we find that the electric field $E=gKg^{-1}$ is constructed with the same $g$ as the vector potential $A=-ig^{\pr}g^{-1}$, where $K=\tilde K \TH(x_0-x_1)$. In addition, $Q$ should be a Hermitian matrix with $Q=g\tilde Q g^{-1}$, where $\tilde Q$ is a constant with respect to $x$, equal to $\tilde K$. In other words $Q$ transform as an $SU(N)$ vector.

It is easy to see how the quantum Gauss' law with sources can be reduced to a classical-like one for the specific functional (\ref{ours}). After simple manipulations we obtain
\be
e^{-i\Omega}G_a e^{i\Omega} \DE(E^2)=-f^k(s)\Gamma_a^k \de(x-x_0)\DE(E^2)+f^k(s)\Gamma_a^k \de(x-x_1)\DE(E^2)
\ee
where $\DE(E^2)$ denotes the delta functions, $\prod_x \de(k_1(x)-\tilde k_1 \TH(x_0,x_1)-\rho_1)... \prod_x \de(k_{N-1}(x)-\tilde k_{N-1} \TH(x_0,x_1)-\rho_{N-1})$, which restrict the amplitude of the electric field. The spinors, $\W$, of the wave functional $\Psi[E]$ have been transfered to the sources, reducing the matrix, $T_a$ to a number. The factor $e^{-i\Omega}G_a e^{i\Omega}$ reduces to a classical-like Gauss' operator giving finally
\be
E^{\prime}_a-f_{abc}E_b{\de \Omega \over \de E_c}=-f^k(s)\Gamma_a^k \de(x-x_0)+f^k(s)\Gamma_a^k \de(x-x_1) \,\, ,
\ee
where $E=gKg^{-1}$ and with the charge density for the two sources being $\bar \rho_a(x)=-f^k(s)\Gamma_a^k$ $ \de(x-x_0)+f^k(s)\Gamma_a^k \de(x-x_1)$. The term ${\de \Omega \over \de E_c}$ is equal to the classical vector potential, $A$, plus another term which contributes to the classical Gauss' operator the additional term $-{1 \over c} \, \tr (K^{\pr} g^{-1}f_{abc}E_b $ ${\partial g \over \partial E_c})$, compensating the non-vector form of the charge density, $\bar \rho_a(x)$. That is, if we combine this term with the charge density $\bar \rho_a$ we obtain $\rho_a$!  The classical configurations are also imposed here by asking the generated vector potential and the electric field to be described with the same $g$ as well as the amplitude to have the form implied by the delta functions, $\DE(E^2)$. 

\section{Conclusions}

In this work we have explicitly solved the Gauss' law for the $SU(N)$ non-Abelian case with static sources in $(1+1)$ dimensions. Two different ways are employed for this task, which result to apparently different solutions; one found from the required transformation properties and the other with explicit evaluation of a proposed ansatz, which resulted to a multiplicity of solutions depending on the representation of the sources. We performed Fourier transform to the first one and showed that it gives the Wilson line solution of the equivalent problem in the A-representation. In addition, we reduced that solution to a superposition of the ones found by using the ansatz, so a complete understanding of the problem is achieved. The Hamiltonian eigenvalues of the wave functionals as well as the connection to the classical case are derived and discussed. The confining potential for the static sources is presented with string tension proportional to the quadratic Casimir operator.

\section{Acknowledgments}

I would like to thank Roman Jackiw and Antonios Tsapalis for helpful conversations.

\end{document}